\documentclass[sigconf]{acmart}

\acmConference[]{}{}{}

\usepackage{bbding}

\setcopyright{none}
\author{Wai Kin Wong}
\affiliation{
  \institution{The Hong Kong University of Science and Technology}
  \country{Hong Kong SAR}
}
\email{wkwongal@cse.ust.hk}

\author[Huaijin Wang]{Huaijin Wang}
\affiliation{
  \institution{The Hong Kong University of Science and Technology}
  \country{Hong Kong SAR}
}
\email{hwangdz@cse.ust.hk}

\author{Zongjie Li}
\affiliation{
  \institution{The Hong Kong University of Science and Technology}
  \country{Hong Kong SAR}
}
\email{zligo@cse.ust.hk}

\author{Zhibo Liu}
\affiliation{
  \institution{The Hong Kong University of Science and Technology}
  \country{Hong Kong SAR}
}
\email{zliudc@cse.ust.hk}
\authornote{Corresponding author.}

\author{Shuai~Wang}
\affiliation{
  \institution{The Hong Kong University of Science and Technology}
  \country{Hong Kong SAR}
}
\email{shuaiw@cse.ust.hk}
\authornotemark[1]

\author{Qiyi Tang}
\affiliation{
  \institution{Tencent Security Keen Lab}
  \country{Shanghai, China}
}
\email{dodgetang@tencent.com}

\author{Sen Nie}
\affiliation{
  \institution{Tencent Security Keen Lab}
  \country{Shanghai, China}
}
\email{snie@tencent.com}

\author{Shi Wu}
\affiliation{
  \institution{Tencent Security Keen Lab}
  \country{Shanghai, China}
}
\email{shiwu@tencent.com}

\renewcommand\footnotetextcopyrightpermission[1]{}

\usepackage{graphicx}
\usepackage{pgfplots}
\usepackage{pgfplotstable}
\usepackage{subcaption}
\usepackage{cellspace, makecell}
\usepackage[many]{tcolorbox}

\usepackage{dsfont}
\usepackage{mathtools,amssymb,latexsym,amsfonts,stmaryrd}
\usepackage{pbox}
\usepackage{xfrac}
\usepackage{booktabs}
\usepackage{xcolor,colortbl}
\usepackage{threeparttable}
\usepackage{amsthm}
\usepackage{scalerel}

\usepackage{hyperref}
\usepackage{multirow}
\usepackage{tikz}
\usepackage{mathrsfs}
\usepackage{mathpartir}
\usepackage[ruled,linesnumbered,vlined,noend]{algorithm2e}
\usepackage{caption}
\usepackage{url}
\usepackage{syntax}
\usepackage{framed}
\usepackage[noend]{algpseudocode}
\usepackage{flushend}
\usepackage{listings}
\usepackage{moresize}
\usepackage{wrapfig}
 
\usepackage{seqsplit}
\usepackage{alltt}
\usepackage{xspace}
\usepackage{enumitem}
\usepackage{pifont}

\DeclareMathAlphabet{\mathcal}{OMS}{cmsy}{m}{n}

\usepackage{amsmath, listings, amsthm, amssymb, proof, xspace}

\usepackage{array}
\newcommand{\PreserveBackslash}[1]{\let\temp=\\#1\let\\=\temp}
\newcolumntype{C}[1]{>{\PreserveBackslash\centering}p{#1}}
\newcolumntype{R}[1]{>{\PreserveBackslash\raggedleft}p{#1}}
\newcolumntype{L}[1]{>{\PreserveBackslash\raggedright}p{#1}}

\usepackage{thmtools}

\declaretheoremstyle[spaceabove=\topsep,notefont=\normalfont\itshape]{mystyle}

\newcommand{\revise}[2]{{\color{red}{\ifx&#1&\else- #1\fi}} {\color{ForestGreen}{\ifx&#2&\else+ #2\fi}}}%
\renewcommand{\revise}[2]{#2}%

\usetikzlibrary{arrows,patterns, decorations.pathreplacing}

\newcommand{\F}{Fig.}

\newcommand{\T}{Table}
\renewcommand{\S}{Sec.}

\newcommand{\ignore}[1]{}

\newcommand{\parh}[1]{\smallskip\noindent\textbf{#1}}

\lstdefinestyle{base}{
  moredelim=**[is][\color{red}]{@}{@},
  escapeinside={<@}{@>}
}

\newcommand{\tool}{\textsc{DecGPT}}

\usepackage{amssymb}
\usepackage{pifont}

\newcommand\DejaVuttfamily{%
  \fontfamily{DejaVuSansMono-TLF}\selectfont }

\lstdefinestyle{base}{
  moredelim=**[is][\color{red}]{@}{@},
  escapeinside={<@}{@>}
}

\lstdefinelanguage
   [x64]{Assembler}     
   [x86masm]{Assembler} 
   {morekeywords={CDQE,CQO,CMPSQ,CMPXCHG16B,JRCXZ,LODSQ,MOVSXD, %
                  POPFQ,PUSHFQ,SCASQ,STOSQ,IRETQ,RDTSCP,SWAPGS, %
                  rax,rdx,rcx,rbx,rsi,rdi,rsp,rbp, %
                  r8,r8d,r8w,r8b,r9,r9d,r9w,r9b}} 

\usepackage[compact]{titlesec}

\usepackage{listings}
\usepackage{fancyvrb}
\usepackage{color}
\definecolor{lightgray}{rgb}{.9,.9,.9}
\definecolor{darkgray}{rgb}{.4,.4,.4}
\definecolor{purple}{rgb}{0.65, 0.12, 0.82}
\definecolor{commentgreen}{RGB}{63,127,95}

\colorlet{myPurple}{blue!40!red}
\definecolor{myOrange}{RGB}{255,192,0}

\newcommand{\enc}[1]{$\phi^{*}_{\theta}$}
\newcommand{\dec}[1]{$\psi^{*}_{\theta}$}

\lstdefinelanguage{Solidity}{
  keywords={len,delete,int,void,payable, public, event, contract, typeof, new, true, false, catch, function, return, null, catch, switch, var, if, in, while, do, else, case, break,struct,const,socklen_t,sa_familty_t,char,sockaddr},
  keywordstyle=\color{violet}\bfseries,
  ndkeywords={class, export, boolean, throw, implements, import, this},
  ndkeywordstyle=\color{darkgray}\bfseries,
  identifierstyle=\color{black},
  sensitive=false,
  comment=[l]{//},
  escapeinside={(*@}{@*)},          
  morecomment=[s]{/*}{*/},
  commentstyle=\color{commentgreen}\ttfamily,
  stringstyle=\color{red}\ttfamily,
  morestring=[b]',
  morestring=[b]"
}

\newcommand{\rnum}[1]{\uppercase\expandafter{\romannumeral #1\relax}}

\usepackage{xcolor}

\algnewcommand{\LeftComment}[1]{\Statex \(\triangleright\) #1}

\definecolor{pptbrown}{RGB}{132,60,12}
\definecolor{pptgreen}{RGB}{56,87,35}

\let\OLDthebibliography\thebibliography
\renewcommand\thebibliography[1]{
  \OLDthebibliography{#1}
  \setlength{\parskip}{0pt}
  \setlength{\itemsep}{0pt plus 0.1ex}
}

\definecolor{pptgreen}{RGB}{84,130,53}
\definecolor{pptred}{RGB}{176,35,24}

\definecolor{pptgreen1}{RGB}{78,173,91}
\definecolor{pptred1}{RGB}{192,0,0}

\definecolor{pptyellow1}{RGB}{203,195,167}
\definecolor{pptgreen2}{RGB}{184,192,176}

\definecolor{pptred3}{RGB}{192,0,0}
\definecolor{pptyellow3}{RGB}{255,192,0}
\definecolor{pptgreen3}{RGB}{4,216,178}

\definecolor{pptblue}{RGB}{0,176,240}
\definecolor{pptgrey}{RGB}{175,171,171}

\newlength{\dpcircle}
\newlength{\rcircle}
\newlength{\dcircle}

\newcommand{\mysubref}[2]{\hyperref[#1]{\ref*{#1}(#2)}}

\newcommand{\mysubRef}[2]{\hyperref[#1]{\ref*{#1}#2}}
\usepackage[normalem]{ulem}

\begin{document}

\title{Refining Decompiled C Code with Large Language Models}

\begin{abstract}

A C decompiler converts an executable (the output from a C compiler) into source
code. The recovered C source code, once re-compiled, is expected to produce an
executable with the same functionality as the original executable. With over
twenty years of development, C decompilers have been widely used in production
to support reverse engineering applications, including legacy software
migration, security retrofitting, software comprehension, and to act as the
first step in launching adversarial software exploitations. Despite the
prosperous development of C decompilers, it is widely acknowledged that
decompiler outputs are mainly used for human consumption, and are not suitable
for automatic recompilation. Often, a substantial amount of \textit{manual
effort} is required to fix the decompiler outputs before they can be recompiled
and executed properly. 

This paper is motived by the recent success of large language models (LLMs) in
comprehending dense corpus of natural language. To alleviate the tedious, costly
and often error-prone manual effort in fixing decompiler outputs, we investigate
the feasibility of using LLMs to augment decompiler outputs, thus delivering
\textit{recompilable decompilation}. Note that different from previous efforts
that focus on augmenting decompiler outputs with higher readability (e.g.,
recovering type/variable names), we focus on augmenting decompiler outputs 
with \textit{recompilability}, meaning to generate code that can be recompiled 
into an executable with the same functionality as the original executable. 

We conduct a pilot study to characterize the obstacles in recompiling the
outputs of the de facto commercial C decompiler --- IDA-Pro. We then propose a
two-step, hybrid approach to augmenting decompiler outputs with LLMs. In
particular, we first launch a static, iterative augmenting step to fix the
syntax errors in the decompiler outputs using LLMs to make it syntactically
``recompilable.'' We then launch a dynamic, memory-error fixing step with LLMs
to fix the memory errors only uncoverable at runtime. The final augmented
decompiler outputs can be smoothly recovered by a C compiler, resulting in a
recompiled executable with the same functionality as the original executable. We
evaluate our approach on a set of popular C test cases, and show that our
approach can deliver a high recompilation success rate to over 75\% with
moderate effort, whereas none of the IDA-Pro's original outputs can be
recompiled. We conclude with a discussion on the limitations of our approach and
promising future research directions.

\end{abstract}

\maketitle

\section{Introduction}

A C decompiler recovers C source code by analyzing and converting the low-level
executable files. Given the pervasive use of C in the software industry, malware
authors, and its unsafe nature, C decompilers have been widely used in software
reverse engineering and security analysis
tasks~\cite{wang2018software,chandramohan2016bingo}. For instance, C decompilers
are often used to recover the source code of legacy software for security
hardening purposes~\cite{david2014tracelet,david2018firmup}. 

To date, many mature C decompilers are available on the market, including
commercial tools like IDA-Pro~\cite{ida} whose licenses cost several thousands
of US dollars, and free ones (e.g., Ghidra) actively maintained by the
open-source community or the National Security Agency (NSA)~\cite{ghidra,nsa}.

Despite the prosperous development and commercialization of C decompilers, it is
widely acknowledged that decompiler outputs are mainly used for human
consumption, and are not suitable for automatic
\textit{recompilation}~\cite{wang2015reassembleable,wang2017ramblr}. Software
compilation is inherently a lossy process, with many high-level information,
such as variable names, type information, and data structures, no longer exists
in the binaries after compilation. Accordingly, decompilers are widely designed
in a pragmatic and conservative manner, where the readability of the generated
code is prioritized over its ``recompilability''~\cite{Liu2020HowFW}, meaning to
generate code that can be recompiled into an executable with the same
functionality as the original executable. 

Despite the fact that decompiler outputs are not suitable for automatic
recompilation, recent research has emphasized the importance of automatically
recompiling the decompiler outputs. For instance, the end goal of various
software cross architecture migration techniques is to recompile the decompiler
outputs into a binary that can run on a different architecture. Also, to reuse
legacy software, it is often necessary to recompile the decompiler outputs into
a binary that can run on a different operating system. More importantly, in
various security instrumentation and hardening tasks, it is often necessary to
instrument the decompiler outputs with security checks and recompile them into a
binary with the same functionality as the original
binary~\cite{dinesh2020retrowrite}. Nevertheless, the progress of enabling
``recompilable'' decompiler outputs has been slow, with the limited work
focusing on rule-based approaches~\cite{Liu2020HowFW} or manual
effort~\cite{mantovani2022convergence}.

To bridge the gap between outputs of de facto C decompilers and the demanding
recompilation requirements, this research first conduct a pilot study to
investigates the challenges faced by recompiling and executing the de facto C
decompiler outputs. We summarize three key challenges by analyzing the 
outputs of the de facto commercial C decompiler --- IDA-Pro. 

Accordingly, we also analyze potential obstacles faced by the state-of-the-art
large language models (LLMs), when being used in an ``out-of-the-box'' setting
to refine the decompiler outputs and deliver recompilable decompilation.

To overcome the obstacles and offer a highly-efficient solution, we propose a
two-step, hybrid framework named \tool\ to augmenting decompiler outputs with
LLMs. First, we employ LLMs to statically fix the syntax errors in the
decompiler outputs. This forms an iterative process, where we feed the augmented
decompiler outputs to the C compiler, and then use the compiler error messages,
if any, to guide the LLMs to re-augment the decompiler outputs in further
iterations. Second, we launch a dynamic, memory-error fixing step with LLMs.
When preparing this step, we compile the augmented decompiler outputs into an
executable with address sanitizer (ASAN)~\cite{serebryany2012as} enabled. We then run the
executable with a set of test cases, and use the ASAN error messages, if any, to
guide the LLMs to fix the memory errors. Our experience shows that a large chunk
of subtle defects can be exposed during runtime by ASAN, and therefore, the
dynamic step effectively exposes the subtle defects that cannot be exposed by
the static step. The final augmented decompiler output will be a piece of C code
that can be smoothly recovered by a C compiler, resulting in a recompiled
executable with the same functionality as the original executable.

We evaluate \tool\ on a subset of Code Contest dataset~\cite{li2022competition},
consisting of 300 test cases, and demonstrate that \tool\ can
significantly increase the recompilation success rate (from the baseline setting
of 45\% to 75\%) with moderate effort. Further ablation study
shows that both the static and dynamic steps are necessary to achieve the high
recompilation success rate. We interpret that the The result indicate generative
AI has promising capabilities in solving fundamental challenges inherent in
reverse engineering. We also discuss several promising future research that can
further enhance the quality of decompilation.

In sum, we make the following contributions:

\begin{itemize}
    \item Conceptually, this study focuses on an important research direction of
    ``recompilable decompilation,'' in response to the increasing demand of
    re-executing decompiler outputs in security and software re-engineering
    tasks. Accordingly, we for the first time explore the feasibility of using
    LLMs to replace previous rule-based or manual efforts to augmenting
    decompiler outputs.
    
    \item Technically, we design a two-step, hybrid approach to deliberately
    unleashing the full potential of LLMs in augmenting decompiler outputs. Our
    approach employs LLMs to statically fix the syntax errors in the decompiler
    outputs, and also dynamically fix the memory errors only uncovered at
    runtime.

    \item Empirically, our evaluation over popular C test cases shows that the
    decompiled C code, once automatically augmented by LLMs, can be smoothly
    recovered by a C compiler, resulting in a recompiled executable with the
    same functionality as the original executable. We also discuss promising
    directions to further improve the quality of decompilation.
\end{itemize}

\parh{Artifact and Data.}~We will release and maintain our artifact after the
paper is officially published.
\section{Background}
\label{sec:background}

To offer a self-contained paper, this section introduces the background of C
decompilers and large language models (LLMs).

\subsection{C Decompilation}
\label{subsec:dec}

\begin{figure}[!htpb]
    \centering
    \includegraphics[width=1\linewidth]{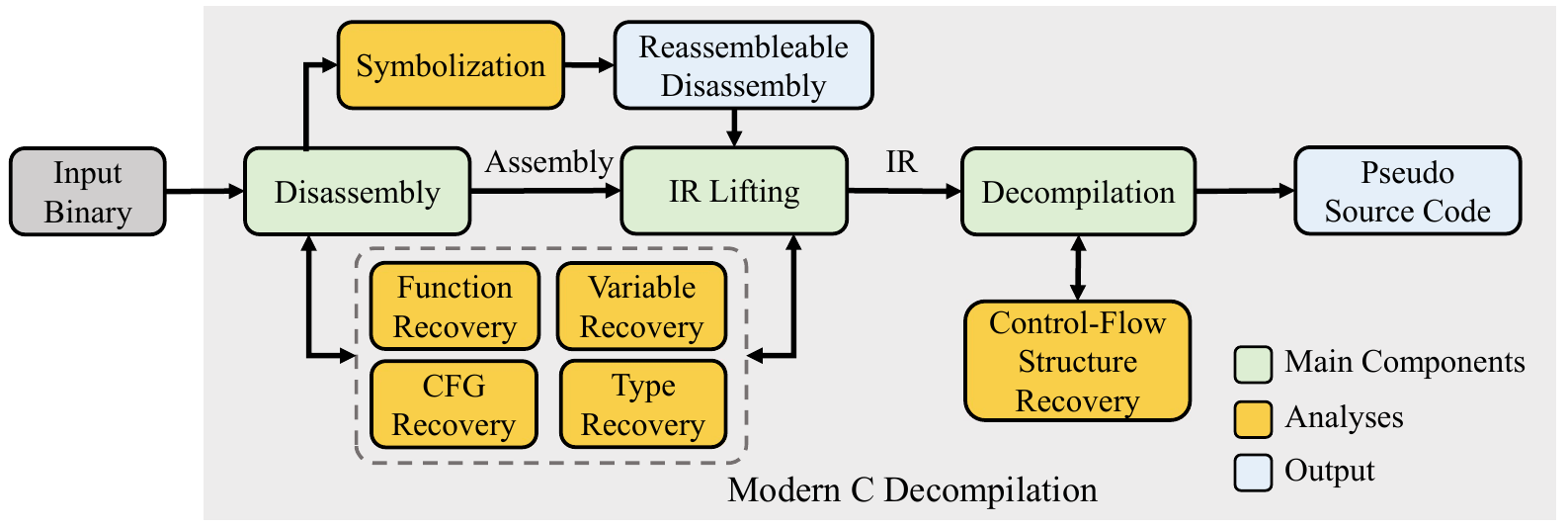}
    \caption{The workflow of modern C decompilers.}
    \label{fig:decompiler}
  \end{figure}

Decompilation refers to the process of reconstructing pseudocode in a high-level
programming language based on low-level assembly instructions extracted from
binary executables. While the concept of decompilation has appeared since the
mid 60s~\cite{halpern1965machine} for the reconstruction of Fortran source code,
the foundation of modern decompilers was proposed in 1994 by
Cifuentes~\cite{cifuentes1994reverse,cifuentes1995decompilation}. Nevertheless,
despite decades of research, modern C decompilation techniques are still not
perfect~\cite{Liu2020HowFW}. The difficulty of decompilation is rooted in the
loss of information during the compilation process. High-level information that
programmers define in source code, e.g., variables, types, and function
prototypes, are discarded by compilers. Compiled binary code only operates
low-level hardware resources like memory and registers. Recovering high-level
abstracts from low-level assembly code is naturally
undecidable~\cite{gurushankar2021minimal}. To solve this problem,
production-level decompilers like IDA Pro and Ghidra typically reconstruct
pseudocode from input binary executable with the following steps. 

\parh{Step1: Disassembly.}~First, the disassembler module of the decompiler
translates the binary into assembly instructions by linearly or recursively
scanning the text section of the
binary~\cite{andriesse2016depth,pang2021sok,pang2022ground}. During the
disassembly process, function boundaries and prototypes are usually identified
with control/data flow analyses~\cite{shin2015recognizing,bao2014byteweight}. 

\parh{Step2: IR Lifting.}~The decompiler will then lift the assembly
instructions into an intermediate representation (IR), which is deemed more
analysis-friendly than assembly code since IR usually contains more high-level
information like types and
variables~\cite{altinay2020binrec,dasgupta2020scalable}. Further analysis will
be applied to the lifted IR for type inference and variable
recovery~\cite{liu2022sok}. While type information does not exist in binary
code, decompilers have to rely on context to infer types. 

\parh{Step3: Code Generation.}~Finally, the decompiler recovers high-level
control flow structures like loops and branches and generates pseudo-source code
as the output~\cite{brumley2013native}. The
decompiler generates control flow graphs (CFGs) based on lifted IR code. Then,
with pre-defined control flow templates, the code generation module will match
against the recovered CFGs and emit the pseudocode structures once a specific
template is matched.

\parh{Challenges.}~As mentioned above, recovering high-level information is
difficult. Specifically, since x86 assembly instructions are not inlined, and
data may be embedded with code, it is non-trivial to distinguish data and
code~\cite{wang2015reassembleable}. The disassembled code is thus usually not
compilable and non-executable unless all symbols (e.g., code and data labels)
are correctly identified. Also, due to complex compiler optimizations, e.g.,
different variables may share the same memory locations, and the limited
scalability of data flow analysis on binary code, variable recovery and type
inference is difficult~\cite{he2018debin,zhang2021osprey,pei2021stateformer}.
Even state-of-the-art commercial decompilers tend to infer the wrong
types~\cite{Liu2020HowFW}, leading to mal-functional and not recompilable
decompilation output. Besides, indirect jump instructions (e.g., \texttt{jmp
rax}) are widely used by C compilers. Such indirect control flow can hardly be
recovered statically, and decompilers may miss part of the CFG or the whole
function~\cite{kim2021refining}. Accordingly, the decompiled code may be broken
and cannot be compiled into a functional binary. Due to the above
challenges, modern decompilers do \textit{not} guarantee
functionality-preserving
decompilation~\cite{burk2022decomperson}, and therefore,
decompiled output usually cannot be used for automatic recompilation, let alone
recompilation with the same functionality as the original executable.

\parh{Recompilable Decompilation.}~We present the following definition of
recompilable decompilation.

\begin{tcolorbox}[size=small]
Given a binary executable $B$, a C decompiler $D$ offers recompilable
decompilation if the decompiled output $O$ can be automatically recompiled by
standard C compilers (e.g., \texttt{gcc}) into a functional binary $B'$ with the
same functionality as $B$.
\end{tcolorbox}

We clarify that recompilable decompilation is largely under-explored due to its
high complexity. Instead, recent works have made promising results on another
relevant pre-step, namely, reassembleable disassembly.
Uroboros~\cite{wang2015reassembleable} first proposed a heuristics-based method
to recover symbols in binaries. Rambler~\cite{wang2017ramblr} extends the
heuristic rules on this basis of Uroboros. Followingly, Superset and
probabilistic disassembly~\cite{bauman2018superset,miller2019probabilistic}
conservatively treat each address as a potential instruction start to make the
disassembled code runnable. The latest solutions,
Retrowrite~\cite{dinesh2020retrowrite} and Egalito~\cite{williams2020egalito},
leverage the auxiliary information presented in position-independent code (PIC)
to achieve recompilable disassembly. However, they cannot be applied to non-PIC
code, and perfect reassembleable disassembly is still not solved for general
executables. 

Recompilable decompilation, on the other hand, greatly improves the user's
ability to modify binary, that is, the user can directly insert high-level
language code in the binary instead of writing assembly code and consequently
can benefit a range of security-related applications, including binary
instrumentation, binary hardening, and legacy software immigration. In
the recompilable decompilation line of research, one recent work leveraging
rule-based methods to fix the decompiler outputs to make them
executable~\cite{Liu2020HowFW}; we present more details in
\S~\ref{sec:motivation}. Similarly, \cite{Reiter2022AutomaticallyMV} performs
partial recompilation by rule based approach for automatic program repairing,
and \cite{mantovani2022convergence} recompiles binaries by manually fixing the
decompilation output. As highlighted by Mantovani et
al.~\cite{mantovani2022convergence}, it takes 90 minutes to 8 hours for
experienced analysts to fix compilation errors when compiling with the
pseudocode, showing the difficulty of recompilable decompilation.

\subsection{Large Language Model (LLM)}
\label{subsec:llm}

LLMs such as GPT-4~\cite{gpt4} and Llama2~\cite{touvron2023llama2} have emerged
as powerful AI assistants, demonstrating remarkable capabilities across diverse
tasks. The massive amounts of training data, as well as techniques such as
Reinforcement Learning from Human Feedback (RLHF)~\cite{christiano2017deeprlhf}
and Direct Preference Optimization (DPO)~\cite{rafailov2023direct}, have enabled
these models to better understand and respond to human instructions, making it
possible to interact with them more naturally. The interaction with LLMs
typically begins with a prompt, which serves as the input to guide the model's
response. Given such an input prompt, an LLM will output a probability
distribution for the next token over its vocabulary. After a token is selected
based on a decoding strategy like beam search or top-k
sampling~\cite{holtzman2019curious}, the new token is appended to the prompt and
used for the next token prediction until a special token indicates the end of
the sequence. 

The performance of LLMs benefits significantly from well-designed prompts that
provide relevant context. It has been shown that task definitions
alone~\cite{liu2021pre} (also known as zero-shot learning) can often achieve
satisfying results, and performance can be further improved by including more
concrete examples in the prompt~\cite{wei2022emergent}. Prompt engineering
techniques have been proposed to improve the performance of LLMs on specific
tasks by carefully designing the input prompt. Among prompt engineering
techniques, Chain-of-Thought (CoT) has been shown to elicit stronger reasoning
from LLMs by asking the model to incorporate intermediate reasoning steps
(rationales) when solving a
problem~\cite{wei2022chain,wang2022rationale,li2022explanations,li2022advance}.

LLMs have manifested their potential and served as ``domain experts'' in a
variety of applications, including code generation, code completion, and code
summarization~\cite{cummins2023compiler,codellama}. In our work, we focus on
leveraging LLMs to augment decompiler outputs with recompilability. We will
present the technical details in \S~\ref{sec:approach}.

\section{Motivation}
\label{sec:motivation}

In this section, we discuss the research motivation behind this work from the
perspectives of research demands, challenges, and technical feasibility.

\subsection{Demanding in Producing ``Recompilable'' C Decompiler Outputs}
\label{subsec:demand}

The proposed approach is designed to augment the decompiler outputs of C
programs to make them recompilable by standard C compilers. The target, namely
``recompilable'' decompiler outputs, is highly demanding in practice. Overall,
various real-world security applications and software re-engineering tasks not
only rely on the decompiler outputs, but also require the automated
recompilation of the decompiler outputs. For instance, in the context of
vulnerability patching, security analysts expect to patch the decompiler outputs
by inserting new and safe code snippets, and then recompile the patched code to
generate a new executable for production usage. Nevertheless, the current
decompiler outputs are not recompilable, and extensive manual effort is usually
required to fix the outputs before they can be recompiled and reused.

We thus argue that the decompiler outputs, whose design goal is to serve human
comprehension, are far from sufficient for many of today's production
applications. This research strives to address the gap by augmenting the
decompiler outputs with LLMs to make them recompilable. 

From another perspective, the proposed approach particularly focuses on the
augmentation of C decompiler outputs. The main reasons are twofold. First, C is
among the most popular and widely used programming languages, and supporting the
augmentation of C decompiler outputs can benefit a large number of real-world,
production scenarios. Moreover, C is an unsafe language, and various security
applications(e.g., vulnerability
detection~\cite{li2018vuldeepecker,chakraborty2021deep},
patching~\cite{Reiter2022AutomaticallyMV,le2012systematic,long2016automatic},
legacy code
hardening~\cite{dinesh2020retrowrite,williams2020egalito,miller2019probabilistic}),
as well as malware analysis~\cite{yan2019classifying,ucci2019survey}, are
particularly targeting C executables.

Second, C decompilations are well-known as a challenging problem in the research
community, whereas recent advances in commercial decompilers like Ghidra and
IDA-Pro have made it possible to generate decompiler outputs with high
functionality accuracy~\cite{Liu2020HowFW}. We thus believe that this study is
timely, addresses an emerging research problem, and can benefit a large number
of real-world applications.

\subsection{Challenges in Recompilation with Outputs from Decompiler}
\label{subsec:challenge}

Following the introduction in \S~\ref{subsec:dec}, we note that prior rule-based
work suffers from various limitations. For example, Liu et al.~\cite{Liu2020HowFW}
enable recompilable decompilation when grammar and types are restricted, yet they
can hardly be applied to general cases. Reiter et al.~\cite{Reiter2022AutomaticallyMV}
circumvented this limitation through partial recompilation combined with binary rewriting,
achieving state-of-the-art results in software hardening. However, this approach does not 
retain a complete copy of recompilable source code.
To understand difficulties in rule-based automatic
recompilation, we randomly selected 100 C/C++ submissions from the Google Code
Jam (GCJ)~\cite{gcj} dataset collected between 2009 and 2020 and investigated the root
causes of failure when recompiling with (pseudo-)source code derived from
decompiler outputs.\footnote{To mimic real-world scenarios where executables are
stripped before releasing, we strip symbols from binaries used in this work. We
also excluded binaries with trivial decompilation errors, which usually are
stack pointer related issues.} 
While decompiler usually supports exporting output as source files, it is mostly
impossible to directly recompile these files due to issues such as undefined
symbols, as outlined in~\cite{Liu2020HowFW}. In order to avoid these issues, the
aforementioned study restricted the number of functions existing in the input
program to one. However, this assumption does not apply to real-world programs,
where it is common for programmers to define multiple functions for solving
challenges. Therefore, we expanded upon the rules used in~\cite{Liu2020HowFW}
and outlined as follows :

\ding{172} In our attempts to recompile the decompilation outputs, we found that
the pseudocode consists of undefined ELF binary-specific symbols, such as
``_gmon_start__'' and ``__cxa_finalize''. These functions are added during
compilation to support the runtime. 

\ding{173} The output from decompilers usually contains security checks
like stack canary checks and array size checks added during compilation. We
removed such checks from the decompiled pseudocode, as they do not impact the
program logic or data structure needed for recompilation. 

\ding{174} Fixing minor errors that affect compilation, such as removing
the ``fastcall'' from the function declaration and fixing the
function type of ``main.''

\ding{175} Fixing the namespace and header based on source code.

\parh{Result.}~After applying the aforementioned rules, only 9 out of the 100 samples can be
recompiled. By executing the recompiled executables and checking the outputs
with the test inputs shipped by the GCJ dataset, we see that all the 9 samples
are functionally equivalent to the original binaries. Nevertheless, the
remaining 91 samples generated a total of 2677 compilation errors. Upon
analyzing these errors, we summarize the following three root causes of
compilation failures:

\ding{172} \underline{Specification error}: This type of error usually happens in the code
generation phases of the decompiler. While the decompilers' signature system may
correctly identify function calls, the final decompiled code does not match API
specifications. A typical error comes from the call towards \texttt{stdin} and
\texttt{stdout}. We observe that decompilers usually break a function call
toward these IO interfaces into several function calls toward other undefined
operators, resulting in undefined function errors when compiling the decompiled
code. Another similar example is the function call towards
\texttt{``std::ios_base::sync_with_stdio.''} While it only consumes a single
boolean type variable, the pseudocode generated from decompilers consists of
other auxiliary variables.

\begin{figure}[t]
    \begin{subfigure}{0.45\textwidth}
    \includegraphics[width=0.87\linewidth]{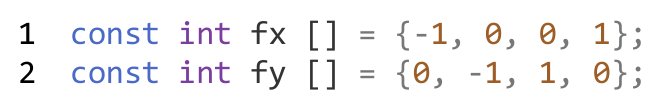}
    \vspace{-5pt}
    \caption{Source code.} 
    \vspace{5pt}
    \label{inference:a}
    \end{subfigure}
    \begin{subfigure}{0.45\textwidth}
    \includegraphics[width=0.99\linewidth]{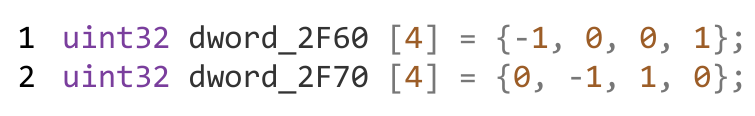}
    \vspace{-5pt}
    \caption{Decompiler's output.}
    \vspace{5pt}
    \label{inference:b}
    \end{subfigure}
    \vspace{-10pt}
    \caption{Example of type inference error.}
    \label{fig:inference}
    \vspace{-0.3cm}
\end{figure}

\ding{173} \underline{Inference error}: During compilation, some information crucial for
humans to understand the program is discarded. Decompilers have to infer such 
information according to the assembly instructions. However, such a process is 
error-prone; decompilers may fail to infer syntactical information 
that is necessary for recompilation. \F~\ref{fig:inference} illustrates 
an example of the type inference error. 

\begin{figure}[t]
    \begin{subfigure}{0.45\textwidth}
    \includegraphics[width=0.87\linewidth]{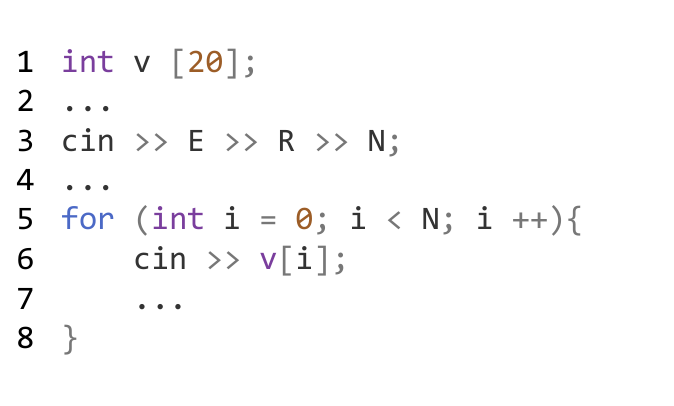}
    \vspace{-15pt}
    \caption{Source code snippets.}
    \vspace{5pt}
    \label{size:a}
    \end{subfigure}
    \begin{subfigure}{0.45\textwidth}
    \includegraphics[width=0.87\linewidth]{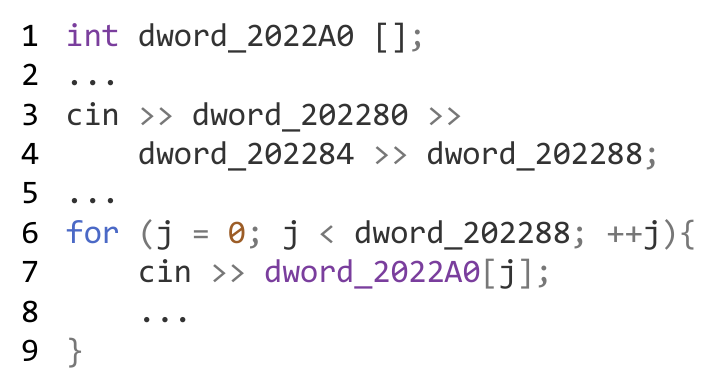}
    \vspace{-5pt}
    \caption{Decompiler's output with minor cleanup.}
    \vspace{5pt}
    \label{size:b}
    \end{subfigure}
    \vspace{-10pt}
    \caption{Example of array size inference error.}
    \label{fig:size}
\end{figure}

As shown in \F~\ref{inference:a}, the actual type of the array is \texttt{const
int}. However, it is inferred as \texttt{uint32} by the decompiler (shown in
\F~\ref{inference:b}). As a result, due to the presence of negative numbers in
the array, it triggered the ``narrowing conversion'' when compiled with default
compilation options. Another common error originates from the inference of array
size. As shown in Fig.~\ref{size:a}, the array size has been hardcoded to 20.
Moreover, we can see the array size cannot be inferred by the decompiler (see
Fig.~\ref{size:b} line 1) and triggered the ``storage size of isn't known''
error. As the array access index \texttt{i} is determined by the user input ``N''
(\texttt{dword_202288} of the decompiled output), the decompiler fails to infer the
size of the array with range analysis. Hence, the array size has been left blank
in the decompilation outputs.

\begin{figure}[t]
    \centering
    \includegraphics[width=0.6\linewidth]{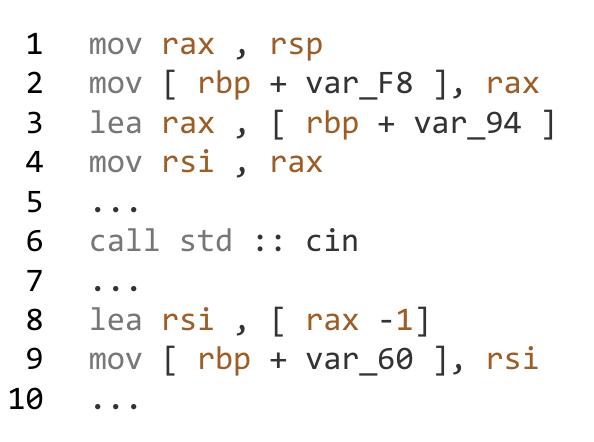}
    \vspace{-5pt}
    \caption{Assembly instruction.}
    \label{failtemplate:c}  
\end{figure}

\begin{figure}[t]
    \begin{subfigure}{0.45\textwidth}
    \includegraphics[width=0.87\linewidth]{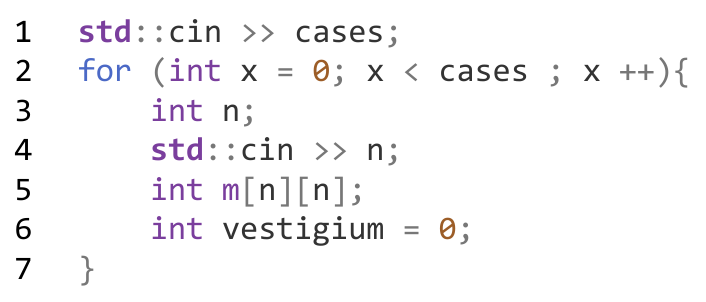}
    \vspace{-5pt}
    \caption{Source code snippets.}
    \vspace{5pt}
    \label{failtemplate:a}
    \end{subfigure}
    \begin{subfigure}{0.45\textwidth}
    \includegraphics[width=0.87\linewidth]{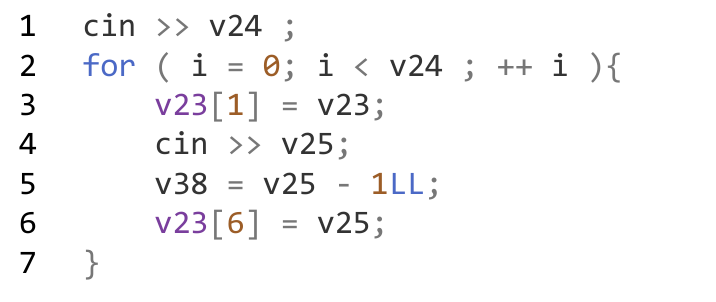}
    \vspace{-5pt}
    \caption{Decompiler's output with minor cleanup.}
    \vspace{5pt}
    \label{failtemplate:b}
    \end{subfigure}
    \vspace{-10pt}
    \caption{Example of decompiler template error.}
    \label{fig:failtemplate}
\end{figure}

\ding{174} \underline{Error from decompiler templates.} As outlined in \S~\ref{subsec:dec},
decompilers' output is generated based on control flow templates. We observe
that some register loading patterns can induce false positives in the template
matching process, resulting in syntactical problems in the generated pseudocode.
We illustrate an example in \F~\ref{fig:failtemplate}, where the decompiler 
generates lines (3, 6) that do not exist in the source code in \F~\ref{failtemplate:a}. 
By examining the corresponding assembly instructions (\F~\ref{failtemplate:c}), we identify 
that the decompiler incorrectly interprets certain stack variable movements (i.e., 
\texttt{mov [rbp+OFFSET], <REGISTER>} in lines 2-9 in \F~\ref{failtemplate:c}) 
as array accesses. Consequently, lines (3, 6) are erroneously generated in 
\F~\ref{failtemplate:b}, leading to invalid conversion errors when recompiling 
with the decompiler's output.

Our manual study shows that specification errors are the most common causes of
recompilation failure, which account for 80.1\% of the total errors. This can be
attributed to the extensive usage of \texttt{stdin} and \texttt{stdout} in
programming contests. Furthermore, although decompilers incorporate signature
databases (e.g., IDA-Pro's Lumina database~\cite{lumina}) to identify function
calls, they often rely on rule-based approaches for the generation of the
decompiled code. However, the presence of diverse combinations of compilers,
optimization passes, and operating systems in real-world scenarios pose a
significant challenge in creating universally-applicable rules that are
completely error-free. It is worth noting that the errors originating from
decompilation templates account for less than 1\% of the total errors,
highlighting the efforts invested by decompiler developers in striving for
near-perfection in their rules.

\subsection{Feasibility and Challenge of using LLMs to Deliver Recompilable Decompilation}
\label{subsec:llmfeasibilitychallenge}

\subsubsection{Feasibility of LLMs.}
\label{subsec:llmfeasibility}

To understand challenges of using LLMs for recompilation, again we performed an
empirical study with 100 randomly sampled programs from the GCJ dataset. We
reported that, the average lines of code and tokens of the dataset is
110.7 and 1082.1 respectively. See \S~\ref{sec:Datasets} for
dataset selection criteria and \S~\ref{subsec:llmselection} for LLMs selection.

With the profound abilities of LLMs, security researchers have been adopting
LLMs to their reverse engineering
pipelines~\cite{gepetto,gptwpre,dalia,xu2023lmpa}. Current LLMs aided reverse
engineering pipelines often adopts a bottoms-up approach to analyzing binary
functions, which has achieve promising results in various tasks like annotating
the decompiler's outputs, predicting function names, and so on. However, this
approach may overlook dependencies across functions, which is necessary for LLMs
to fix inference error by interpret code-level semantics. Therefore, we include
the complete outputs from decompiler in each query to the LLM. We interacted
with ChatGPT in this way until $N$ iterations or successfully recompiling the
test cases. $N$ is configured as 15 per our prelminary study, and we present
other configurations at this step in \S~\ref{subsec:modesetting}.\footnote{As a
fair setup, we do not obtain information like headers or namespace from source
code. Instead, we use LLM to infer this.}

Notably, previous works have demonstrated that the provided context can significantly influence 
the output generated by LLMs. Methods such as few-shot learning~\cite{wei2022emergent} 
and Chain-of-Thought (CoT)~\cite{wei2022chain} have been proposed to improve performance on 
complex reasoning tasks. However, these techniques are not directly applicable to our recompilation 
task, since they typically require specialized context samples to be supplied, which is difficult to 
obtain for general recompilation scenarios. Therefore, to maintain reproducibility while preserving 
effectiveness, inspired by prior works on using LLMs for automated program repair (APR)~\cite{Xia2023ConversationalAP}
we designed a conversational approach with a carefully designed system prompt
(shown in \T~\ref{tab:prompt1}) to guide the model to correct the decompiled
outputs iteratively. The zero-shot method avoids reliance on specific training
samples, while the system prompt is crafted to focus the model on the
recompilation task. This allows our approach to be applied in a generalizable
manner across different decompilation outputs.

To be specific, the system prompt is designed with the three considerations.
First, we provide explicit requirements for the generated outputs to ensure GPT
do not generate outputs for another programming languages nor using Windows C
convention, for example using \texttt{wmain} as program's entry point. Second,
we re-emphasize the scope to GPT, to avoid the model neglecting the function
entry point (i.e. \texttt{main}) and results in compilation failures. Third, we
clearly stated the dos and don'ts for recompilation to the model. This avoid the
model to only return explanations or directly copy the ``goto'' syntax from the
decompiler outputs, without understanding the semantics. For user prompt, we
formatted the decompiled pseudocode with inline code formatting in Markdown.
This is a well-known prompt engineering technique~\cite{promptengineeringcourse,
openaicookbook} to indicate our pseudocode from the inputted prompts and
providing cues to the expected outputs from the model.  

For handling compilation errors and output errors, we use the same system prompt
with extra messages appended before the problematic code, as outlined in
\T~\ref{tab:prompt1}. 

\begin{table*}[htbp]
    \centering  
    \caption{Prompt templates designed for our preliminary studies.}
    \label{tab:prompt1}  
    \resizebox{0.90\linewidth}{!}{
    \begin{tabular}{|m{2cm}|m{12cm}<{\raggedright}|m{10cm}<{\raggedright}|}
        \hline
         & Prompt \\
        \hline
        System Prompt & 
        ``\textit{Generate linux compilable C/C++ code of the main 
        and other functions in the supplied snippet without using goto, fix any missing headers.
        Do not explain anything.}''
        \\
        \hline
        Compilation Error & 
        ``\textit{Please fix the following compilation errors in the source code:  \{compiler_error\} \{pseudocode\}}''
        \\
        \hline
        Output Error & ``\textit{The expected output of the program for input: \{expected_input\} is \{expected_output\}, 
        but we got \{wrong_output\}. Please fix the issue in the source code: \{pseudocode\}}''
        \\
        \hline
    \end{tabular}
    }
\end{table*}

\subsubsection{Challenges of LLMs.}
\label{subsec:llmchallenge}

With aforementioned settings, 42 out of the 100 samples can recompile
successfully. To understand the remaining obstacles, we manually analyzed all
failed samples and identified the three root causes of failure: 

\parh{Limited Context Length.}~Despite explicit instructions to ChatGPT,
it occasionally ``forget'' the instructions given by our input prompts. This can
manifest in providing explanations of functionalities or returning mal-formed
pseudocode, etc. Overall, context length is a common weaknesses of LLMs in solving 
long inputs~\cite{brown2020language, kaddour2023challenges}. It frequently happens 
regardless of the configurations or the training methodologies of the
model~\cite{llmforget,liu2023lost,longchat}.

\parh{Repairing with Shortcuts.}~Shortcut learning
behavior~\cite{geirhos2020shortcut} refers to a phenomenon where a decision
performs well on benchmarks but fails to generalize or address the underlying
problems. This issue has been observed across various learning-related
tasks~\cite{chowdhery2022palm,dagaev2023too,pezeshki2021gradient}. Upon
analyzing failure samples, we have identified similar issues in ChatGPT.
Specifically, when encountering type-related errors, ChatGPT tends to utilize
casting operators like ``static_cast'' or ``reinterpret_cast'' to wrap
statements. While this approach may appear reasonable initially, it often
results in memory corruption during testing. Additionally, ChatGPT may attempt
to fix compilation errors by removing related code fragments or functions. While
this may seem like a reasonable action to eliminate compilation errors, it leads
to the loss of contextual information in the decompiler outputs and ultimately
results in recompilation failure. This behavior mirrors a similar issue observed
in the field of Automated Program Repair (APR)~\cite{Qi2015AnAO}, where
plausible patches generated by APR tools overfit towards specific test suites
used for patch synthesis, failing to address the root cause of defects. These
observations highlight the tendency of LLMs, similar to APR, to generate patches
that are tailored specifically to the provided conditions, potentially limiting
the generalizability and robustness of the recompilation process.

\begin{figure}[t]
    \begin{subfigure}{0.45\textwidth}
    \begin{lstlisting}[basicstyle=\scriptsize\ttfamily]
    v14 &= byte_2022A0[2048 * m + n];
    \end{lstlisting}
    \vspace{-5pt}
    \caption{ChatGPT's output.} 
    \vspace{5pt}
    \label{hallucination:a}
    \end{subfigure}
    \begin{subfigure}{0.45\textwidth}
    \begin{lstlisting}[basicstyle=\scriptsize\ttfamily]
    v14 |= byte_2022A0[2048 * m + n];
    \end{lstlisting}
    \vspace{-5pt}
    \caption{Decompiler's output.}
    \vspace{5pt}
    \label{hallucination:b}
    \end{subfigure}
    \caption{Example of hallucination error.}
    \label{fig:hallucination}
\end{figure}

\parh{Hallucination.}~Typically, hallucination refers to LLMs generating content
that deviates from real-world facts learned during
training~\cite{hallucination2023}. Although factuality may not be the primary
objective of generated code, similar phoenomenon is observed in the
decompilation process, wherein critical semantics of the code are wrongly
altered by LLMs. Specifically, when a long sequence of code tokens are fed to
the LLM, the model may return the results that slightly change the original code
tokens. An interesting example of such hallucination is shown in
Fig.~\ref{fig:hallucination}, where the bitwise operator \texttt{\&=} has been
replaced with \texttt{|=} after two conversations with LLMs. Since the variable
\texttt{v14} is used as a status flag in the following conditional statement,
this minor defect results in a significant change in the data flow, leading to
wrong answer returns from the recompiled program.

\subsection{Distinction with Previous Works Augmenting Decompiled Code}

Below discussing the technical details, we first clarify the following key
differences between our work and previous works in related fields. In short, our
focus is to using LLMs to fix the decompilation errors such that the decompiled
code can be smoothly recompiled by standard C/C++ compilers, whose output is
functionally equivalent to the original binaries. In other words, we do not aim
to ``beautify'' the decompiled code, but to make it automated recompilable. We
have clarified the demanding of recompilation in \S~\ref{subsec:demand}.

In contrast, another line of research aims at augmenting decompiled code to make
it more readable. Often, such works do not directly improve the result
correctness, but try to recover various high-level program information and
enhance the readability for humans, e.g., an anti-malware expert can read the
enhanced decompiled code more easily if the variable names are precisely
recovered and very meaningly.

To this end, DIRE~\cite{lacomis2019dire} and DIRTY~\cite{chen2022augmenting} try
to improve the readability of decompiled code by assigning decompiled variables
meaningful names and types with NLP models. Nero~\cite{david2020neural}, on the
other hand, generates function names for decompiled code. Similarly,
NFRE~\cite{gao2021lightweight} and SymLM~\cite{jin2022symlm} try to predict
function names for stripped binaries. Such works do not improve the result
correctness but ease humans and LLM to comprehend. Moreover,
Debin~\cite{he2018debin} and CATI~\cite{chen2020cati} predict debugging
information from binaries without decompilation. A recent work,
LmPA~\cite{xu2023lmpa} combines program analysis with LLM to predict names of
decompiled functions. As aforementioned, this paper is \textit{orthogonal} with
the above works, as we study the automatic recompilation of decompiled code into
functional output, which is rarely studied in the literature except those few
papers listed in the \textbf{Recompilable Decompilation} paragraph of
\S~\ref{sec:background}.

\begin{figure}[!htpb]
    \centering
    \includegraphics[width=1\linewidth]{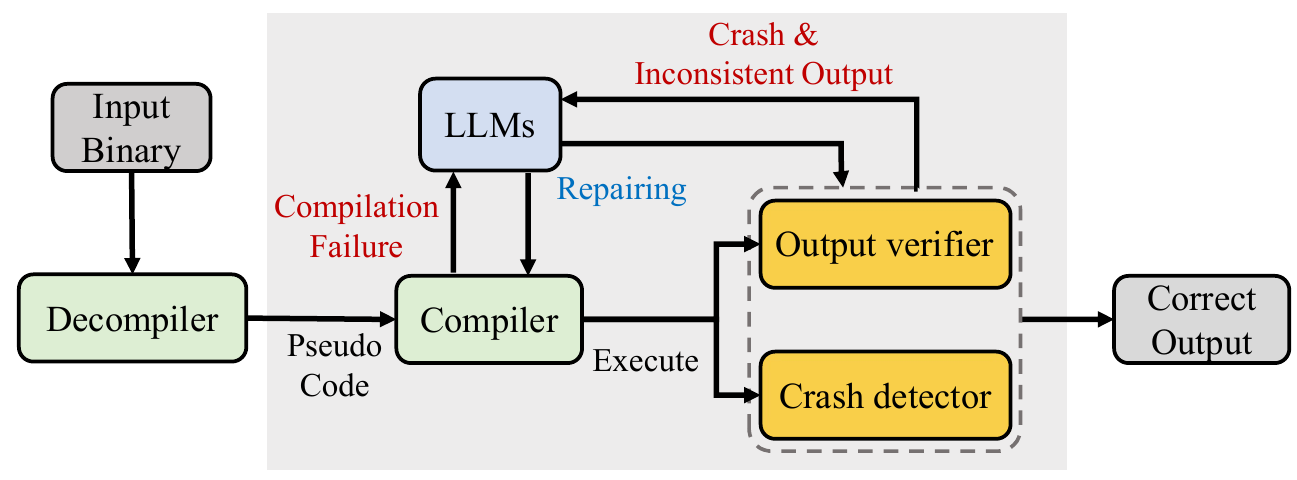}
    \caption{The workflow of \tool.}
    \label{fig:workflow}
  \end{figure}

\section{\tool: Enabling Automated Recompilable Decompilation with LLMs}
\label{sec:approach}

In line with the preliminary studies presented in \S~\ref{sec:motivation}, we
see the promising technical potential of LLMs in assisting the recompilation
process. LLMs behave as ``human experts'' to handle the tedious, costly and
error-prone manual effort in fixing decompiler outputs. Given that said,
\S~\ref{subsec:llmfeasibilitychallenge} has illustrated that such LLM-based
approach is by no means trivial. In this section, we present our approach,
\tool, that offers a two-step, hybrid pipeline to augment decompiler outputs
with LLMs.

\F~\ref{fig:workflow} depicts the workflow of \tool. We first launch a static,
iterative augmenting step to fix decompiler outputs and make them compilable
by standard C compilers (details in \S~\ref{subsec:static}). Then, we launch a dynamic,
memory-error fixing step to fix the memory errors detected during runtime
(see \S~\ref{subsec:dynamic}). The final augmented decompiler outputs can be
smoothly recovered by C compilers to generate a semantics correct executable.
Before we dive into the details of our approach, we first introduce the
application scopes and other setups below. 

\subsection{Application Scopes and Setup}
\label{subsec:llmselection}

\begin{table*}[!htbp]
  \caption{Designed prompt templates for \tool.}
  \label{tab:prompt2}  
  \centering  
  \resizebox{0.9\linewidth}{!}{

  \begin{tabular}{|m{2cm}|m{12cm}<{\raggedright}|m{10cm}<{\raggedright}|}
    \hline
       & Prompt \\
      \hline
      System Prompt & 
      ``\textit{Generate linux compilable C++ code of the main and other functions in the supplied 
      snippet without using goto, fix any missing headers and reducing the number of 
      intermediate variable. Only reply the fixed source code. Do not explain anything 
      and include any extra instructions, only print the fixed source code.}''
      \\
      \hline
      Compilation Error & 
      ``\textit{Please fix the following compilation errors in the source code:  \{compiler_error\} \{pseudocode\}}''
      \\
      \hline
      Output Error & ``\textit{The expected output of the program for input: \{expected_input\} is \{expected_output\}, 
      but we got \{wrong_output\}. Please fix the issue in the source code: \{pseudocode\}}''
      \\
      \hline
      ASAN Error & 
      ``\textit{Please fix the \{type_of_memory_corruption\} triggered in \{statement\}:  \{pseudocode\}}''
      \\
      \hline
  \end{tabular}
  }
\end{table*}

\parh{Decompiler Selection.}~The designed \tool\ aims to augment C/C++
decompiler outputs with LLMs to make it re-compilable by standard C/C++
compilers. We have clarified the design focus on C/C++ decompilers in
\S~\ref{subsec:demand}. Given that said, our technical pipeline is not limited
to C/C++ decompilers. We believe that our approach can be easily extended to
decompilers for different programming languages; we present further discussions
in \S~\ref{sec:discussion}. In this study, we focus on the de facto commercial C
decompiler --- IDA-Pro~\cite{ida}. IDA-Pro is an extensively used C/C++
decompiler that domains both industrial and academic usages.

\parh{LLM Selection and Setup.}~In this study, we investigate the potential of
LLMs as a black box for assisting the recompilation process. We specifically
focus on GPT-3.5~\cite{gpt35}, a readily accessible LLM model, and interact with
it automatically through its API, ``gpt-3.5-turbo-0613''. Similar to its
predecessor, GPT-3~\cite{brown2020language}, GPT-3.5 undergoes pre-training on a
large corpus of web data and subsequent fine-tuning using Reinforcement Learning
from Human Feedback (RLHF)\cite{christiano2017deeprlhf}, as outlined in
\S~\ref{subsec:llm}. This enables the model to generate responses to user
queries. 

Although GPT-4~\cite{gpt4}, the successor of GPT-3.5, is considered to be the
most powerful LLM available in the wild, GPT-3.5 still offers several practical
advantages. First, in terms of the cost of querying the model, our observation
shows that GPT-3.5 is about 20 times more cost-effective based on current
pricing structures~\cite{openaiprice}. Second, GPT-3.5 provides at least 2 times
faster response~\cite{gpt4time} than GPT-4. Therefore, GPT-3.5 is a more
suitable and cost-effective LLM for our study. Given that said, our technical
pipeline is not limited to a certain LLM, and it is possible to use other LLM
models as replacements. We present further discussions in
\S~\ref{sec:discussion}.

\subsection{\tool --- Static Augmenting}
\label{subsec:static}

The static augmenting step aims to fix the grammatical errors and inference
errors in the decompiler outputs and make the outputs compilable by standard
C/C++ compilers. Given a piece of decompiler output $o$, we first compile the
output with a standard compiler $C$ (in our implemention, we use the standard
GCC compiler). If $C$ fails to compile $o$, we then launch the following static
augmenting step.

\parh{\ding{172} Initial Prompting.}~The static augmenting step is an
iterative process. With default decompiler outputs, we apply 
preprocessing rules \ding{172} to \ding{174} as documented in 
\S~\ref{subsec:llmchallenge}. This removes unnecessary tokens from the pseudocode and 
helps prevent hallucination originated from these. Expanded upon the 
settings in \S~\ref{subsec:llmfeasibility}, we instruct the model 
to reduce redundant variable assignments found in the decompiler outputs (as 
illustrated in \T~\ref{tab:prompt2}). 
By doing this, we are striving to reduce output length and also the chance 
of hallucination if the source snippet is required for further repairing 
with LLM. With the above steps, we prepare an initial prompt for LLM to fix the inputs. 

\parh{\ding{173} Post Processing.}~Although we specified the desired output
format in our prompts, our testing has revealed that ChatGPT sometimes produces
partially mal-formed outputs that include code explanations. Using these outputs
directly for recompilation often leads to compilation errors. To address this
issue, we create a set of rules to cleanup the output $o$. After processing, we
re-compile $o$ with the standard C/C++ compiler.

\parh{\ding{174} Processing Error Message.}~If $o$ fails to be compiled, we
collect the compiler error messages $E$ during the compilation. Then, we first
preprocess $E$ to remove the irrelevant information, e.g., the line numbers and
file names, with scripts. We then feed the preprocessed error messages $E$ with
certain lines that consist of the errors into the LLM to generate a new token
sequence $T$. 

\parh{\ding{175} Repairing Decompiled Output $o$.}~At this step, we then
tokenize the decompiler output $o$ into a sequence of tokens $T$. Then, we
prepare a prompt $P$ to be fed into the LLM. The prompt $P$ instructs the LLM to
fix the token sequence $T$ and generate a new token sequence $T'$. 

\parh{\ding{176} Iterative Augmenting.}~The fixed token sequence $T'$ is then
fed into the compiler $C$ to re-compile the repaired output $o'$. If $C$ again
fails to compile $o'$, we firstly iterate over each function in $o'$  and check
if certain function's function body is being stripped; if so, we revert the
changes. We then repeat the above process from \ding{172} until $C$ successfully
compiles $o'$, or the number of iterations exceeds a pre-defined threshold $N$.
In our current implementation, we set $N$ to 15 based on our preliminary
experiments. 

The above static augmenting step aims to fix the grammatical errors and
inference errors in the decompiler outputs and make the outputs compilable by
standard C/C++ compilers. However, the outputs may still contain various memory
errors and functional errors. In other words, even if the output of this static
repairing phase is ``re-compilable'', its induced executable may still crash or
produce obviously incorrect results. To address this issue, we launch the
following dynamic repairing step.

\subsection{\tool --- Dynamic Repairing}
\label{subsec:dynamic}

\parh{Motivation and Design Consideration.}~The dynamic repairing step aims to
fix the functional errors in the decompiler outputs. While this appears to be a
straightforward and demanding task, our tentative exploration shows that
precisely pinpointing and fixing various functionality errors in the decompiler
outputs is a very challenging task, whose technical solutions are still in their
infancy. Note that performing automated bug detection in C code is very
challenging, let alone we are dealing with decompiler outputs. Similarly,
automatically ``patching'' the detected bugs in C code is also a very
challenging task. While the community has made some encouraging progress on
LLM-based bug detection and patching, our preliminary study shows that these
approaches are not directly applicable to our problem.

Thus, this research takes a pragmatic approach. Instead of somehow instructing
the LLMs to search and fix arbitrary functional errors in the decompiler
outputs, we focus on fixing the memory errors in the decompiler outputs. Memory
errors are particularly common in C programs, and are often the root cause of
functional errors. Moreover, the C/C++ community has developed full-fledged
tools to detect and fix memory errors, with the help of memory address
sanitizers (ASAN)~\cite{serebryany2012as}. As a result, we target on leveraging
the existing tools to detecting memory errors in the decompiler outputs, and
then instructing the LLMs to fix the detected memory errors. We aim to present
such meaningful benchmarking results to the community, and leave exploring
technical solutions to fix arbitrary functional errors in the decompiler outputs
as future work. 

\parh{Approach.}~At this step, we profile the compiled executable $E$ with a set
of test cases $T$. Moreover, we configure the C compiler to inject address
sanitizers into $E$ during the compilation stage. This way, whenever $E$
contains subtle memory errors, the address sanitizers shall faithfully detect
and report them during the runtime profiling phase. With the memory detection
results, we then launch the following dynamic repairing step. 

\parh{\ding{177} Collecting Memory Error Information.}~We first collect the
memory error information $I$ when one of the address sanitizer injected in $E$
alarms on the program input $t$. The memory error information $I$ includes the
address, the erroneous instruction, the register values in the context, and the
stack trace. This information is well-formed by address sanitizers, and can be
easily parsed for analysis using scripts. 

\parh{\ding{176} Repairing Defects.}~We then prepare a prompt $P$ to be fed into
the LLM. The prompt $P$ instructs the LLM to fix the memory error information
$I$ originated in the test case $t$. The LLM then generates a new token sequence
$T'$.

\parh{\ding{176} Testing Functionality Equivalence.}~If test case $t$ does not
result in an alarm, we will verify the output of $E$ against the expected
output. If the output is correct, we proceed to the next test case. If the
output is incorrect, we prepare a prompt $P$ and instruct the LLM to fix the
functional defects \textit{at our best effort}; again, we believe directly
fixing functional defects is a very challenging task, and our main focus is on
the above memory error fixing. Overall, we then repeat the above process from
step \ding{177} until $E$ passes all the test cases or fails any test cases in
$T$, and instruct the LLM to fix the identified defects.
\section{Evaluation Setup}
\label{sec:Experimental}

As previous discussed in \S~\ref{subsec:llmselection}, we implement \tool\ for
supporting the decompilation outputs for IDA-Pro~\cite{ida}, a commercial and
well tested decompiler. \tool\ is primarily written in Python and C++,
with about 2504 lines of code. All experiment are conducted on a Ryzen 3970X
32-core server with 256GB memory. Below, we introduce the evaluation setup.

\subsection{Model Setting}
\label{subsec:modesetting}

We limited our selection of programs based on the maximum token length
limit of our meployed GPT-3.5 model, which is 4096 for the GPT version
``gpt-3.5-turbo-0613''. Notably, this limit includes both the input and output
parts of the program. As such, to ensure that the model had enough capacity to
generate outputs, we only selected programs where the total number of tokens in
the decompiled pseudocode and the system prompt was less than half of the input
token length limit. This is because the length of the recompiled code is roughly
equivalent to the length of the input decompiled pseudocode, and our system
prompt had fewer tokens. By doing this, we were able to generate complete
recompiled programs and avoid any errors that may have been caused by
truncation, which could have affected the experimental results.

\subsection{Datasets}
\label{sec:Datasets}

In our evaluation, we use the Code Contest dataset~\cite{li2022competition}, which
is a diverse collection of submissions obtained from five online judge platforms. Each
problem in the dataset consists of multiple test cases used to validate the correctness
of the submissions, with the "AC" (Accepted) verdict indicating that all test cases were
passed successfully. In addition, we specifically selected submissions with AC verdicts
for our evaluation as it indicates that the corresponding code submissions have successfully 
passed all the provided test cases. This enables us to assess the correctness of the recompiled 
binaries based on the supplied test cases, and ensure the reliability of the evaluation results.

To address the uncertainty about the source code in the training set of ChatGPT, we took 
precautions in our evaluation. We only considered submissions made after September 2021, which 
is the knowledge cutoff of ChatGPT~\cite{openaimodeloverview}. Also, context length plays a crucial
role in the performance of LLMs. To account for this, We divided the dataset into 
five equal intervals based on the context length and randomly selected 60 programs from each interval. 
This gave us a dataset of 300 programs. We manually verified that there is no trivial decompilation errors
presence in the decompilation outputs, and we reported that the average lines of code and tokens 
is 112.9 and 1110.4, respectively.

\subsection{Evaluation Metrics}

Since the recompilation process enabled by \tool\ is an iterative process, we
use the length of the conversation chain (denoted as $C$) required to correctly
recompile decompiler's output as our primary evaluation metric. Due to the cost,
we evaluated with $C$ = 1, 5, 10 and 15. We deem a success if the recompiled
binaries passed all test cases, without triggering errors.

\section{Evaluation}
\label{sec:Evaluation}

\parh{Research Questions.}~Our evaluation aims to answer the following research questions:

\parh{RQ1}: How effective is \tool\ in solving the challenges of 
rule-based recompilation? 

\parh{RQ2}: How effective is \tool\ in alleviating the challenges of 
LLM-based recompilation? 

\parh{RQ3}: How much does the iterative design contribute to the 
performance of \tool.

\begin{table}[h]
  \caption{Recompilation success rate for 3 different tools.}
  \label{tab:rq1}
  \centering
  \resizebox{0.8\linewidth}{!}{
    \begin{tabular}{c|c|c|c}
      \hline
       & \tool\ & \textsc{DecRule} & $llm-baseline$ \\
      \hline
      \textbf{Success Rate $C$ = 1 }   &37\% &  8\% &  32\%   \\
      \textbf{Success Rate $C$ = 5 }   &55\% &  8\% &  39\%   \\
      \textbf{Success Rate $C$ = 10}   &69\% &  8\% &  42\%   \\
      \textbf{Success Rate $C$ = 15}   &75\% &  8\% &  45\%   \\
      \hline
    \end{tabular}
  }
\end{table}

\subsection{RQ1: Comparison with Rule-Based Recompilation}
\label{sec:rq1}

To evaluate the effectiveness of \tool\ in dealing with the challenges of
rule-based recompilation, we compared it with the settings we used in
\S~\ref{subsec:challenge}; to ease presentation, we denote the setting used in
\S~\ref{subsec:challenge} as \textsc{DecRule}. To ensure a fair comparison with
\tool, we added the header and namespace extracted from the original source code
to the decompilation results. Note that in the case of \tool, these information
is expected to be inferred by LLM.

As shown in \T~\ref{tab:rq1}, \textsc{DecRule} can successfully recompile
8\% of binaries from the testing dataset, while \tool\ can recompile
75\% of them. This comparison highlights the effectiveness of \tool\ in
addressing the challenges associated with rule-based recompilation.

To understand the types of errors that \tool\ is capable of fixing, we use the
compilation errors generated when using \textsc{DecRule} as a reference and
manually examine the conversation chain between the LLM and \tool. Our analysis
reveals that about 98\% of the specification errors are being fixed by \tool\
within two conversation rounds, regardless of whether the test case can be
recompiled or not. We believe the findings are reasonable. To generically fix
specification errors, heavy-weighted binary analyses like symbolic execution are
usually employed to infer the function prototype~\cite{choi2021ntfuzz}. In
contrast, LLMs can fix these errors relatively easily by replacing the
syntactically incorrect code block with the correct one. This is possible
because typical APIs and function prototypes are widely represented in the
training sets used for training the LLMs or can be inferred from the compilation
errors submitted to the LLMs.

With our design, \tool\ can fix 62.5\% of inference error and all of the 
identified decompiler template errors. By leveraging decompiled pseudocode as input 
and using compilation errors as feedback, \tool\ iteratively replace erroneous segments 
within the decompiler outputs. 
In one of the successful recompiled cases, we found that the initially
decompiled output contains a type inference error, where a variable's integer
type is incorrectly inferred as an unsigned integer type. Subsequently, this
variable and another correctly inferred variable are used to invoke the
\texttt{max} function. While this results in compilation errors, we find that
with two iterations of conversation, \tool\ can successfully fix this issue by
precisely replacing the unsigned integer type to the integer type. Similar
situation also happens for LLM to fix the decompiler template errors, which
usually results in type conversion failure during compilation.

However, \tool\ may not be able to fix all of the inference errors. Out of the
remaining 37.5\% of samples, we found that 70.2\% of them come from the
Standard Template Library (STL) functions. STL is an important part of the C++
programming paradigm and offers reusable containers and algorithms. Note that
the identification of STL functions is generally considered
difficult~\cite{igorstl}, in the sense that decompiled code of the same STL
function with different types can be significantly different. In our study, we observed that the presence
of STL function calls tends to increase the context length of the decompiler outputs from 2.8 times 
to 16.1 times. This poses a challenge for LLMs, as the performance of LLMs is highly dependent on the
context length~\cite{liu2023lost}. However, we also view this as a challenge from decompilations,
where the decompiled pseudocode may not accurately represent the used STL functions, thereby
hindering LLMs in fixing syntactical errors.

The remaining errors all come from ill-inferred buffer sizes in different
contexts. While buffer size can be indirectly fixed through ASAN error messages,
lacking of accurate buffer size information still posing challenges for LLMs to
successfully recompile the binaries. We consider this as a challenge for both
LLMs and the decompiler, due to the fact that the performance of LLMs highly
depends on the quality of its inputs (i.e., decompiler outputs).

\begin{table}[h]
  \caption{Number of success cases in terms of varying context lengths across different settings.}
  \label{tab:rq2}
  \centering
  \resizebox{0.8\linewidth}{!}{
    \begin{tabular}{c|c|c|c}
      \hline
       & \tool\ & $llm-baseline$ & \tool\ Zero \\
      \hline
      \textbf{200 - 560}     &57	 & 48 & 58\\
      \textbf{560 - 920}     &54	 & 45 & 50\\
      \textbf{920 - 1280}    &48	 & 24 & 10\\
      \textbf{1280 - 1640}   &39	 & 9  & 5\\
      \textbf{1640 - 2048}   &27	 & 6  & 2\\
      \hline
    \end{tabular}
  }
\end{table}

\subsection{RQ2: Comparison with LLM-Based Recompilation}
\label{sec:rq2}

We now evaluate \tool\ to understand its effectiveness in addressing the challenges 
of LLM-based recompilation that we identify in \S~\ref{subsec:llmchallenge} 
(referred to as $llm-baseline$). Compare to $llm-baseline$, \tool\ has made advancements 
in several aspects. First, \tool\ extends the system prompt for enhancing the overall 
stability of the recompilation process. Second, a structural level sanity check is added
to the source code returned by LLM, and ensure the input to the model in next iteration includes all the necessary 
information for recompiling a binary. Third, we compile the binaries with ASAN enabled.
This allows us to extract the line of statement that triggers the memory corruption when 
binary crashes. We then use this information as part of the prompt for LLM to generate a 
version of the output that fixes the problems. Tables \ref{tab:rq1} and 
\ref{tab:rq2} show the success rate and the number of successful cases per interval for these 
two settings. Overall, \tool\ outperforming $llm-baseline$ by 30\% in terms of success rate in 
recompiling the same set of binaries; it is seen as more capable in handling decompiler outputs with
long context. We interpret the results as follows.

\parh{Effectiveness of the Extended System Prompt.}~To evaluate the effect of
the extended system prompt in \tool, we refer to the result of $C$ = 1 in
\T~\ref{tab:rq1}. In this case, the success rate is measured on binaries
successful recompiled without extra repairing query. Compare to the result of
$llm-baseline$, it is easy to observe an immediate increase in the success rate
from 32\% to 37\%, when the system prompt is extended. In other words, we
interpret that by extending the system prompt, \tool\ effectively imposes more
constraints on the LLM and helps the LLM to concentrate on fixing decompiler
outputs and aligns more closely with the desired output. These contribute to the
improved performance without additional iterations.

\parh{Effectiveness of Static Augmenting.}~\tool\ outperforms $llm-baseline$
when recompiling outputs with long context lengths. As shown in
\T~\ref{tab:rq2}, \tool\ successfully re-compiles 4.3 times and 4.5 times more
decompiler outputs than that of $llm-baseline$ for context lengths between 1280
to 1640 and 1640 to 2048, respectively. This suggests that \tool\ can effectively
alleviate the challenges of LLM-based recompilation for longer context lengths. 

This notable difference in performance is attributed to the synergy effect of
extended system prompt and static augmenting. When examining the success cases
within the two aforementioned context length intervals, we found that \tool\ is
capable of detecting and reverting a total of 194 instances where LLM recklessly
strips function bodies during the fixing process. \tool\ effectively preserves
the integrity of the decompiler outputs by avoiding these undesirable
modifications, and ensures that necessary context can pass to the next iteration
of fixing. As a result, \tool\ demonstrates better performance than
$llm-baseline$ in recompilation with long context inputs.

\parh{Effectiveness of Dynamic Augmenting.}~The design of dynamic augmenting is
to address functional incorrectness that cannot be resolved through static
augmenting alone. To evaluate the success of fixes provided by the dynamic
augmenting of \tool, we manually reviewed 50 cases where \tool\ successfully
fixed a runtime errors, and we categorized these fixes into three categories.
The two major categories are output formatting errors and minor algorithmic
errors, which accounted for 48\% and 42\% of the sampled cases, respectively.
Output formatting errors typically involve issues related to the structure or
formatting of the output, like missing a newline after each output. As for minor
algorithmic errors, they typically include small mistakes such as a reversed
plus/minus sign or reversed program logic. These problems can be fixed by LLMs
when the correct output is provided as part of the user prompt. The remaining
10\% of the fixed cases belongs to memory corruption issues that stem from
inference errors in the decompiler outputs.

\parh{Limitations of Dynamic Augmenting.}~An important part of the dynamic
augmenting is the use of ASAN to detect memory corruption issues. Out of the 55
ASAN instances triggered during runtime, \tool\ is able to fix 40\% of them. We
believe the results as reasonable and generally encouraging, showing the
potential of incorporating ASAN (whose outputs are documents of severe memory
errors in a well-structured format) with LLM to augment the decompiled outputs.

We further investigate the cause of the unfixable ASAN errors by manually examine
the intermediate conversation between LLM and \tool. We categorized the cause of
unfixable ASAN into two types. The major cause of unfixable ASAN errors belongs to 
the usage of type casting operators (like ``static_cast'' and ``reinterpret_cast'') when 
LLM attempted to fix type conversion errors. It contributes to 81.8\% of the unfixable ASAN errors. 
Despite attempts to explicitly disallow the usage of these type casting operators in the system prompt or user 
prompt, LLM continued to employ them as the fixing strategy. This behavior may be attributed to the shortcut 
learning behavior discussed in \S~\ref{subsec:llmchallenge}. However, the underlying reasons 
and appropriate mitigations for this behavior are left as future work, as they are beyond the 
scope of the current paper. The remaining 18.2\% of unfixable ASAN errors are caused by errors 
associated with hallucination, like wrong buffer size being infered by LLM, as we discussed in
\S~\ref{sec:rq1}.

\subsection{RQ3: Ablation Study}
\label{sec:rq3}

To understand the contribution of the iterative and two-phase design to the
performance of \tool, we run \tool\ without either the static augmenting or
dynamic augmenting mentioned in \S~\ref{sec:approach}, and we denote this
setting as \tool\ Zero. Similar to \tool, we use this setting to query LLM 15
times, and we consider the recompilation a success if any of the outputs
returned from LLM passes all of the test cases (using the same criteria as
\tool). 

The results are shown in \T~\ref{tab:rq2}. Generally, \tool\ has clear advantage
over \tool\ Zero, especially for handling decompiler outputs with long context.
For the context length between 200 to 560 and 560 to 920, we observe a
comparable performance between \tool\ and \tool\ Zero. The result is reasonable,
given that the average number of iterations requires for \tool\ to fix the
decompiler output in these two intervals are 1.24 and 2.13, respectively. With
such a small number of iterations, the performance of \tool\ and \tool\ Zero are
similar.

However, for the remaining intervals, \tool\ Zero shows a notably lower success
rate than the iterative design. This is reasonable: with an increase in context
length, more functions are present in the decompiler outputs, which implies that
more syntactical errors may be present in the decompiler outputs as well.
Another technical obstacle for LLMs in handling long context is the performance
degradation~\cite{liu2023lost} and information forgetting
problems~\cite{llmforget}. With the iterative design, \tool\ can fix the errors
in a step-by-step manner by providing sufficient context for LLMs to resolve
existing syntactical errors present in decompiler outputs or errors introduced
by LLMs in previous iterations. In contrast, \tool\ Zero only has one chance to
fix the errors, and as a result, \tool\ Zero is more likely to suffer from the
obstacles mentioned above. These results suggest that the iterative design is a
key factor in the performance of \tool.

\section{Discussion}
\label{sec:discussion}

\parh{Extension to Other Decompilers.}~\tool\ is evaluated to augment the
outputs of IDA Pro, the de facto commercial C/C++ decompiler that is extensively
used in industry and research. Nevertheless, we believe the proposed approach is
independent of the underlying C/C++ decompiler. As illustrated in
\S~\ref{sec:approach}, in the static augmenting stage, we leverage error
messages from the compiler to refine the decompilation output. Consequently,
\tool\ requires the decompiler to output pseudo source code that is close to the
source code consumed by compilers. To our observation, this prerequisite is
consistent with the concept of software decompilation, and is generally
satisfied by mainstream decompilers that are available on the market. Besides,
given that the incredible power of LLM is not limited to the output of any
specific decompiler, we envision that LLM is able to read and understand the
decompilation output of other mainstream C/C++ decompilers, and also decompilers
of other programming languages. \tool, therefore, is expected to be extended to
different decompilers without showing a significant difference in effectiveness.

\parh{Validity of Using The Current LLM.}~One potential threat to the validity
of our study is whether \tool\ can be applied to other LLMs. To ensure
generality, we deliberately refrain from conducting specific hyperparameter
tuning in our research. Instead, we evaluate \tool\ using the default
configuration of GPT-3.5. This eases the reproducibility of our work and
enhances the generalizability of our findings. Additionally, we conduct a
comprehensive evaluation of each component of \tool\ in \S~\ref{sec:Evaluation}
to assess their effectiveness. Based on this analysis, it should be accurate to
conclude that \tool\ is not limited to specific LLMs. Furthermore, we perform
our empirical study (in \S~\ref{sec:motivation}) and evaluation (in
\S~\ref{sec:Evaluation}) on two separate datasets, both of which demonstrate
comparable performance. This indicates that the performance of \tool\ is not
restricted to a specific dataset.
\section{Related Work}
\label{sec:related}

We have reviewed the general workflow of software decompilation and recent
progress in decompilation, particularly on recompilation, in
\S~\ref{sec:background}. Below, we review other research works that are relevant
to this paper.

\parh{Automatic Program Repair (APR).}~The goal of APR is to fix defeats in
software automatically. Generate-and-validate is a typical approach of APR by
synthesizing patches base on the fault localization results. A patch will be
consider as plausible patch if it passes all of the test case and fix the issue. 
In short, patches are usually synthesized by three main approaches, namely
search-based approach~\cite{Goues2012GenProgAG,long2015staged}, template-based
approach~\cite{kim2013automatic} and semantic-based
approach~\cite{li2020dlfix,long2016automatic}. By leveraging LLMs's capabilities
in suggesting code snippets base on user's prompts~\cite{lemieux2023codamosa},
recent study~\cite{Xia2023ConversationalAP,xia2023automated,
fan2023automated,jiang2023impact} has demonstrated encouraging results in
applying LLMs to APR. In short, by carefully expressing localized defective code
fragment or failure as prompts, it is shown that LLMs generates potential
patches through conversational prompting. In contrast, this study employs LLMs
in a hybrid pipeline to fix decompiler outputs, a well-known challenging problem
in reverse engineering that has many distinct characteristics from APR of source
code.

\parh{Dynamic Binary Rewriting.}~Binary rewriting refers to the process of
changing the semantics of a program executable file without access to its source
code. Most of its applications are in the field of software security, such as
defeating code reuse attack, software obfuscation, and software watermarking.
While we mainly discuss static binary rewriting in \S~\ref{sec:background},
dynamic rewriting techniques have also developed significantly in the past few
decades. Dynamic rewriting techniques are performed during the runtime, often on
the basis of system-level support like Intel Pin~\cite{luk2005pin},
DynamoRIO~\cite{bruening2003infrastructure}, and
Valgrind~\cite{nethercote2007valgrind}. Those tools intercept program execution
at the runtime with an external or grafted process. Dynamic binary rewriting
advantages itself by enabling modification of all code without restrictions
suffered by static methods~\cite{duck2020binary,dinesh2020retrowrite}. However,
it incurs a high overhead because the cost of rewriting occurs at runtime. In
contrast, this work's focus lies in static binary rewriting, which is ahead of
program execution and is more efficient. The success of our proposed approach
shall further extend the capability of static binary rewriting methods, i.e.,
enabling smoothly instrumenting a piece of binary code with modest cost as how
source code is easily modified.

\parh{Large Language Model for Code}~In recent years, LLMs have made significant
strides in enhancing various code-related tasks. Commercial models such as
Github Copilot~\cite{copilot} and OpenAI GPT-4~\cite{gpt4} have demonstrated
impressive performance, while several open-source models have been developed
using large-scale code corpora.

Incoder~\cite{DBLP:journals/corr/abs-2204-05999}, for instance, employs a causal
masking training objective to excel in code infilling and synthesis, while
CodeGen~\cite{nijkamp2022codegen} is a pre-trained model for multi-turn program
synthesis with over 16B parameters. The BigCode Project has also contributed to
the development of StarCoder~\cite{DBLP:journals/corr/abs-2305-06161}, an
open-source model with 15.5B parameters. More recently,
Code-Llama~\cite{codellama}, a code-specialized version of Llama 2, was created
by further training on code-specific datasets and sampling more data from the
same dataset for longer context. Apart from LLMs that focus on source-code level
tasks, models have been developed for low-level code. For example,
CodeCMR~\cite{yu2020codecmr} and IRGEN~\cite{li2022unleashing} are pre-trained
models designed for low-level code on various code-related tasks. Additionally,
researchers~\cite{cummins2023compiler} have attempted to train a 7B-parameter
model from scratch on low-level code to optimize LLVM assembly for code size.
Overall, these LLMs have demonstrated impressive performance in various
code-related tasks, with potential for further advancements in the field.

\section{Conclusion}

This research is motivated by the high demand of delivering recompilable
decompilation in security and software re-engineering tasks. Moreover, we are
also motivated by the recent major success of LLMs in comprehending dense corpus
of both natural language text and programs. To bridge the gap between outputs of
de facto commercial C/C++ decompilers and the demanding recompilation
requirements, we propose a two-step, hybrid framework named \tool\ to augmenting
decompiler outputs with LLMs. Evaluations show that \tool\ can significantly
increase the recompilation success rate with moderate effort. We conclude with a
discussion on promising future research directions on top of \tool.

\bibliographystyle{ACM-Reference-Format}
\bibliography{bib/main,bib/similarity,bib/ref,bib/decompiler,bib/cv,bib/llm,bib/zj,bib/apr}

\end{document}